# Limits on Relief through Constrained Exchange on Random Graphs

Randall A. LaViolette[*], Lory A. Ellebracht[†] and Charles J. Gieseler[‡]

**(revised)**

## Abstract

Agents are represented by nodes on a random graph (e.g., "small world"). Each agent is endowed with a zero-mean random value that may be either positive or negative. All agents attempt to find relief, i.e., to reduce the magnitude of that initial value, to zero if possible, through exchanges. The exchange occurs only between the agents that are linked, a constraint that turns out to dominate the results. The exchange process continues until Pareto equilibrium is achieved. Only 40%-90% of the agents achieved relief on small world graphs with mean degree between 2 and 40. Even fewer agents achieved relief on scale-free-like graphs with a truncated power-law degree distribution. The rate at which relief grew with increasing degree was slow, only at most logarithmic for all of the graphs considered; viewed in reverse, the fraction of nodes that achieve relief is resilient to the removal of links.

[*] Sandia National Laboratories, Albuquerque NM 87185 USA ralavio@sandia.gov
[†] Sandia National Laboratories, Albuquerque NM 87185 USA lcooper@sandia.gov
[‡] RESPEC, 5971 Jefferson NE Suite 101, Albuquerque NM 87109 USA cjgiese@sandia.gov

**Introduction**

We describe a simple model of a system of agents who each create excesses internally and independently of other agents. The agents are required to relieve their excess through exchanges with other agents. Relief is the state of zero excess because surpluses and deficits are equally undesirable for the agents. Agents interact only when their own excesses drive them to seek relief through exchanges. We represent agents by nodes on a graph, where the links represent a route for a feasible exchange between pairs of agents. The exchange between two directly linked agents occurs if and only if one of the agents possesses a surplus and the other a deficit. Here we are interested in exchanges between agents on graphs of moderate degree that may be qualitatively representative of agent interactions in a wide variety of physical or social contexts [1-6]. Restricting the number of other agents that can exchange with a given agent turns out to impose strong limits on the relief that can be achieved even when (within these constraints) the best possible exchanges occur. The following section provides a detailed description of the model. Subsequent sections provide results, discussion, and conclusions.

In order to give this exchange process a more formal economic interpretation, the agent's excess may be imagined to be one of two goods, one good represented numerically as a positive quantity (surplus) while the other good represented as a negative quantity (deficit). Each agent has a Leontief (fixed proportions) utility function such that utility gains are only possible when equal quantities of goods are obtained [7]. With this representation, the supply and demand curves are perfectly inelastic; the quantity demanded or supplied is independent of price, which is set to unity for the whole system. An agent maximizes its utility when it is able to balance its quantity of one good with an equal quantity of the other good. Because each agent is given an initial endowment, an agent is able to achieve a higher utility than his initial state by trading, but is restricted in his trading partners to only its neighbors, as determined by the graph. The exchange process continues until one of the many possible Pareto equilibria is achieved [7].

This study is related to but nevertheless different from studies that have been concerned with wealth distributions achieved through exchanges of goods. In some of these studies, the assumption that all agents interact with all other agents seems at least implicit [8-11], in network terms, this corresponds to placing agents on the complete graph. There has been some recent work with agents exchanging on sparse graphs [12, 13]; the agents trade with traditional (exponential) utility functions in order to establish a market price for the exchange. A related annihilation model also was studied on small world graphs [14]. We did not address price formation or wealth distributions in this work. Instead we postpone to future investigation the gains from exchange for individual agents. The social efficiency of exchanges on networks was another question we did not address but this has been studied with a detailed exchange model that employed auctions between heterogeneous agents on bipartite subgraphs [15]. The model also resembles some aspects of the AB-percolation model but we did not employ the results of AB-percolation here[16].

## Description of the model

### Constructing the Graphs

All of the graphs employed here were simply connected with undirected links and were constructed with some kind of random process except for the square lattice. The complete graph was excluded from this study as we directed our attention to graphs of moderate degree. Trees (acyclic graphs) also were excluded from this study. All graphs were constructed with a minimum degree of two so that no agent would be dependent on only one other for relief. Most of the results presented below were for the small world graph, which has been widely employed in recent studies of abstract social systems [1]. We employed the variant of the small world graph in which extra links were randomly added to a graph initially composed of one, two or three offset rings (denoted by "$r = 1$", "$r = 2$", and "$r = 3$" below, with minimum node degree of two, four and six respectively), that contained all of the nodes, as employed by, e.g., Bollobás *et al.* [17] or Guclu *et al.* [18] (see also [19-21]) instead of the rewiring algorithm in the original Watts-Strogatz model [22]. A few realizations of the Erdös-Rényi (E-R) random graph [23] were also employed for comparison. One instance of a 2025 node ($45 \times 45$) square lattice, without periodic boundary conditions, was also included. Finally, a few realizations of a truncated power-law graph were constructed with the "random configuration model" [1, 24] so that each of the scale-free-like graphs were generated with their density distribution of degree $k$ proportional to $\exp(-k/c)/k$ for $2 \le k \le 160$ and with $c$ chosen to provide the various mean degrees.

The number of nodes was set to $N = 2000$ for all graphs except for the 2025-node square lattice. The mean degree is $2E/N$ where $E$ is the number of links. In constructing the graphs from the various algorithms, only the largest connected component (i.e., the giant component) of the graph was chosen; in practice this precaution had a negligible effect on the results.

### Initializing the Agents

Each of the $N$ nodes (agents) of the graph was assigned an "excess", i.e., a real number $x$ drawn randomly from a real-valued density distribution $g$, symmetric about zero, with zero mean and median. The excess is a "surplus" if $x$ is positive and a "deficit" if $x$ is negative (alternatively, it is one good or the other corresponding to the sign of $x$). For this study we chose $g$ to be a slightly modified normal distribution with standard deviation of $\sigma = 10$; the mean magnitude of the initial excess was $\sigma\sqrt{2/\pi} \approx 7.98$. The modifications consisted of ensuring that the magnitude of every excess was at least 0.001 (accomplished by redrawing the excess otherwise) and subtracting from it the mean excess of that sample (much less than 0.001 in magnitude) so that each resulting sample of the excesses summed to zero. This initialization process produced a small fraction of nodes ineligible to participate in the exchange because all of their neighbors were initialized with the same sign of excess.

## Exchange Process

After initializing each node's respective excess, the exchange process was initiated by uniformly randomly selecting a link. An exchange would take place between its nodes (say *j* and *k*) if and only if $x_j \cdot x_k < 0$, for which case the new values $x'_j, x'_k$ would become

$$x'_j = 0$$

$$x'_k = x_j + x_k$$

for $|x_j| \leq |x_k|$.

With one of the node's excess set to 0 after the exchange, no further exchanges can occur on that link, so it is removed from the list (but not the graph) of eligible links. The link selection continued until no links remained eligible for exchanges, i.e., $x_j \cdot x_k \geq 0$ for all *j* and *k* that shared a link. The final configuration is one of many Pareto equilibria; alternatively it is one of the many absorbing states of this random walk. The link selection process excludes next-nearest neighbor exchanges between nodes.

A node's excess either remains constant or becomes reduced in magnitude after an exchange. In each exchange the node of lesser magnitude would be relieved and would become ineligible for further exchanges. The other node nevertheless receives a reduction of its excess and remains eligible for further exchanges. No agent moves farther from relief as the result of any exchange. No node acquires more of the opposite excess than is needed for it to achieve relief, i.e., the excess may become zero but it never switches sign.

The resulting excesses (including the zero excesses of the relieved nodes) contributed one sample to the distribution of excess. Multiple samples were obtained by resetting all excess values on the same graph and repeating the exchange procedure above 100 times in order to eliminate any dependence of the result on the order of the link selection. Multiple realizations were achieved by generating 100 graphs of each type, for each mean degree.

## Results

For each graph, the distribution of excess at the end of the exchange process was symmetric and nearly normal, also with zero mean but with a smaller variance. Figure 1 displays the mean of the fraction of nodes relieved ($f_{relieved}$) on a semi-log plot, for various graphs with varying mean degree. Figure 1 also displays the mean of the fraction of nodes ($f_{eligible}$) initially eligible to participate in an exchange. The density distributions of $f_{relieved}$ and $f_{eligible}$ were each essentially normal for each graph, with one deviation displayed as error bars in Figure 1; in some cases, the deviation is smaller than the size of the symbol in the plot. The semi-log plot in Figure 1 indicates that the mean $f_{relieved}$ increased at most logarithmically with increasing degree, for any kind of graph. For each graph, except for the square lattice, the mean $f_{relieved}$ was well above its respective site percolation threshold [25]; for the square lattice it was just below its site percolation threshold (0.55 vs. 0.59). Nevertheless the relief was far from complete for any graph, even for those graphs in which nearly all nodes were initially eligible to participate in an exchange. For the E-R and small world graphs, a mean degree of about 40 was required to achieve even 90% relief, for which the degree distributions of the small world graphs had become essentially the same as those of the E-R graph (i.e., Poisson, see [23]). Much less relief was achieved in the truncated power-law (scale-free-like) graphs compared to the small world and E-R random graphs.

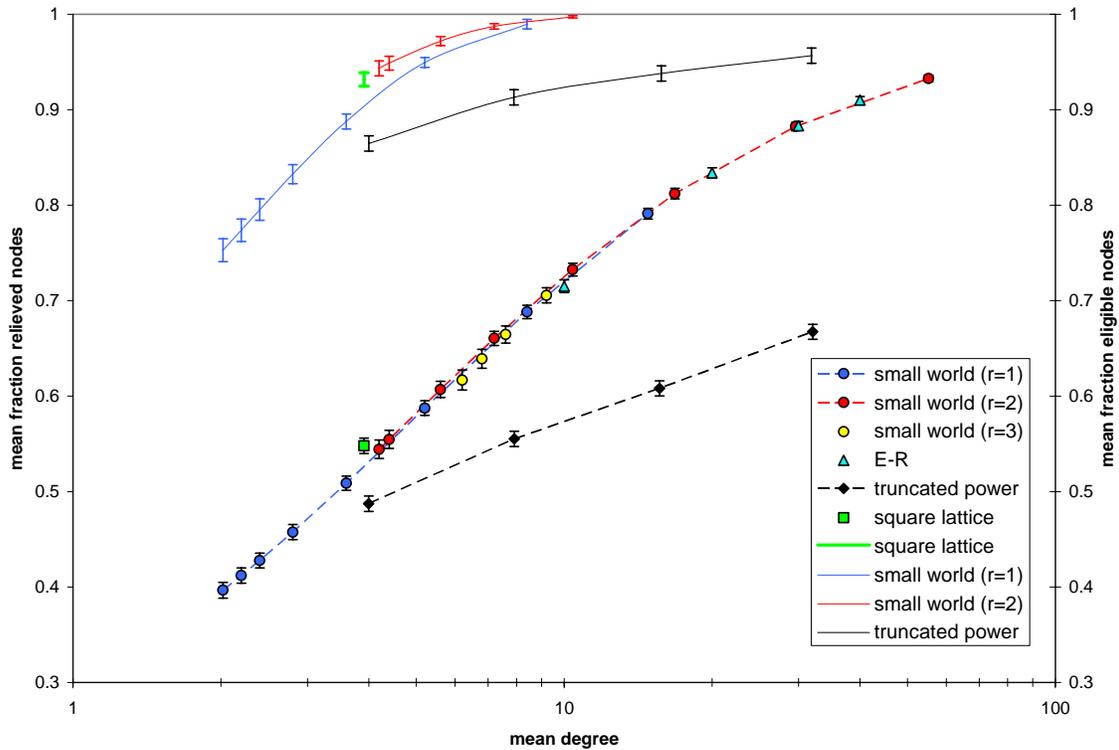

**Figure 1. The mean fraction of relieved nodes and the mean fraction of eligible nodes vs. the mean degree for various graphs.** The symbols, some with dashed straight lines (see legend), are the values for the fraction of relieved nodes. The solid curves are splines between the values for the fraction eligible nodes. The "error bars" correspond to uncertainty intervals of plus or minus one standard deviation, respectively.

Figure 2 shows a semi-log plot of typical results for the fraction of relieved nodes as a function of the node's degree, which indicates a slowly increasing frequency of relief for the nodes of increasing degree. Figure 2 shows only the results for which the standard deviation is small; where the number of nodes of a given degree is small the uncertainty becomes large.

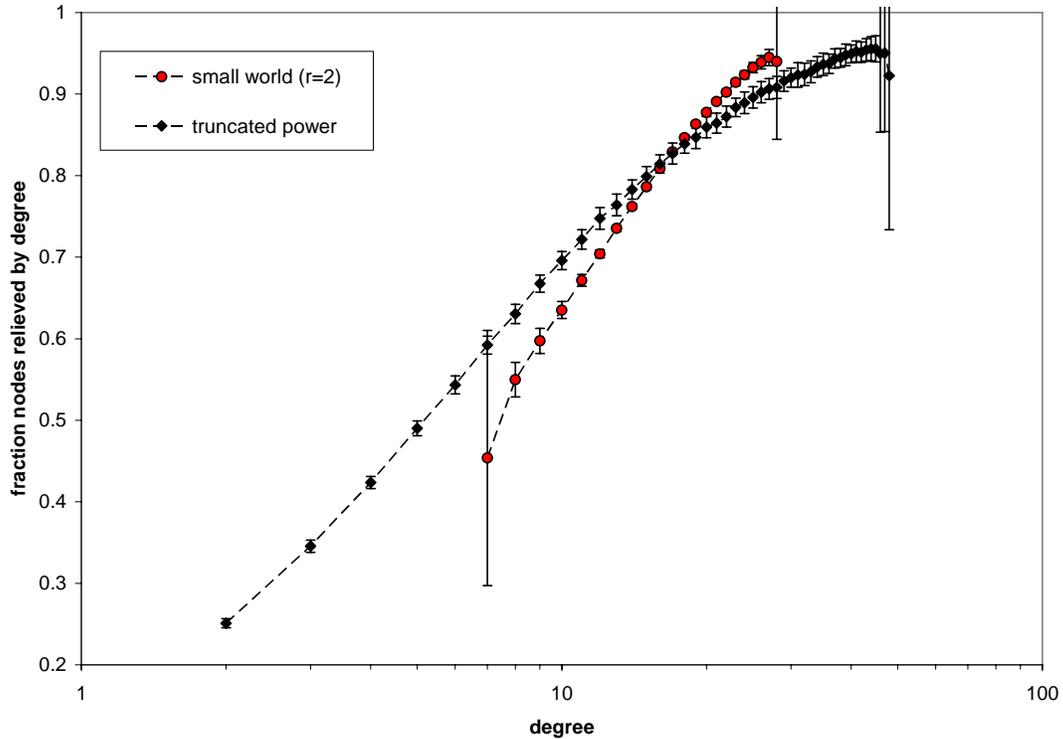

**Figure 2. Mean fraction of nodes relieved by degree for two comparable graphs.** The mean degree of the small world (r = 2) and the truncated power law graph was 16.8 and 15.8, respectively. The solid "error bars" correspond to uncertainty intervals of plus or minus one standard deviation, respectively. The results were truncated where the standard deviation grew rapidly, resulting from a small population of nodes with that degree. The dashed lines are guides to the eye.

We did not show the degree distribution of the relieved nodes for each graph because it always closely tracked the degree distribution of the original graph. The results were insensitive to a wide range of possible choices for the initial parameters, distributions and rules for the exchange. In particular, the results were insensitive to increasing the number of trials and increasing the number of graphs that contribute to the samples. They were also insensitive to increasing the number of nodes in the graphs. The results were insensitive to the differences in the disorder between the different kinds of small world graphs, between the small world and the E-R random graphs, and even between the small world or E-R graphs and the square lattice. Variations in the initial distribution of the excess had a negligible impact on the results, as long as the distribution was continuous and symmetric about zero, a remark that applies not only to the choice of the deviation for the normal distribution, but even on the kind of distribution (normal vs. bounded uniform). We initially thought that forcing the exchanges to be incremental by placing limits on the magnitude of each exchange would allow more nodes to achieve relief. Nevertheless, for limits as small as 0.1, the results obtained (for the graphs mentioned in Figure 2) with these limits were within a standard deviation of those obtained without limits.

We did not pursue here an examination of the stability of the metastable equilibria achieved by the exchange process but we note that the following sequence of perturbations slowly leads to a system without any relieved agents: assign new excesses only to all of the relieved agents and reapply the exchange process until a new equilibrium is achieved. At least a thousand cycles were required to eliminate relieved agents. We defer a discussion of this evolution to future work.

## Discussion and Conclusions

This simple model has a complicated solution, without a closed form (as far as we know) for any but the complete graph [26] yet it lacks the usual indicators of complexity or "system behavior" [1-3]. The distribution of excess at the end of the process is essentially normal even on the truncated power-law graphs. The model does not display any kind of phase transition. The model shows no statistically significant difference in the equilibrium that results from exchanges that proceed for the maximum benefit of the pair of agents involved and those which resulted from exchanges that were limited by an external rule.

The difference in relief achieved by the scale-free-like truncated power law graphs on the one hand and the small world and E-R graphs on the other suggests that, for the same mean degree, the graph with the more homogeneous degree distribution achieves more relief and that the benefits of homogeneity increase with increasing mean degree. The truncated power law graphs have a few highly connected nodes that always get relief but a large number of low degree nodes that get much less relief; in other words, those few highly connected nodes that frequently got relief contributed very little to the relief of the whole system. This conclusion is not surprising in view of similar conclusions reached in recent work on transport and synchronization efficiency on complex networks [27-30].

Complete relief should be achieved with this exchange process on a complete graph but we were surprised by the high numbers of links that were required to achieve even 90% relief on the much less connected graphs studied here. We were not surprised that increasing the mean degree of the graph increased the relief and that the higher degree nodes received more relief than the lower degree nodes but we were surprised by the slow, at most logarithmic growth in relief with increasing degree. If the cost of acquiring a link to another agent remains constant (e.g., no discounts), then the cumulative costs of acquiring links would grow linearly with degree. Viewed in reverse, the results suggest that the achievement of relief is resilient (or insensitive) to node or edge removal for all graphs.

## Acknowledgements

We thank Eli Ben-Naim (Los Alamos National Laboratory), Paul Parris (University of Missouri at Rolla), Robert McGraw (Brookhaven National Laboratory) and Sidney Redner (Boston University) for helpful discussions. The work was supported in part by the Sandia National Laboratories Laboratory Directed Research and Development program and the National Infrastructure and Simulation Analysis Center. The National

Infrastructure Simulation and Analysis Center is joint program at Sandia National Laboratories and Los Alamos National Laboratory, funded by the United States Department of Homeland Security's Preparedness Directorate, under the direction of the Infrastructure Protection/Risk Management Division.  Sandia National Laboratories is a multi-program laboratory operated by Sandia Corporation, a Lockheed Martin Company for the United States Department of Energy's National Nuclear Security Administration under contract DE-AC04-94AL85000.